\documentclass{jps-cp}
\usepackage{txfonts} 

\usepackage{graphicx}
\usepackage[abs]{overpic} 
\usepackage{color}
\usepackage{longtable}
\usepackage{upgreek}
\usepackage{tikz}

\title{Current Results on Light Hypernuclei in the WASA-FRS HypHI Experiment}

\author{
  C.~\textsc{Rappold}$^{1}$, 
  P.~\textsc{Achenbach}$^{2,3}$ , 
  H.~\textsc{Alibrahim Alfaki}$^{4}$ , 
  F.~\textsc{Amjad}$^{4}$ , 
  M.~\textsc{Armstrong}$^{4,5}$ , 
  K.-H.~\textsc{Behr}$^{4}$ , 
  J.~\textsc{Benlliure}$^{6}$ , 
  Z.~\textsc{Brencic}$^{7,8}$ , 
  D.~\textsc{Calongue González}$^{1}$ , 
  T.~\textsc{Dickel}$^{4,9}$ , 
  V.~\textsc{Drozd}$^{4,10}$ , 
  S.~\textsc{Dubey}$^{4}$ , 
  H.~\textsc{Ekawa}$^{11}$ , 
  S.~\textsc{Escrig}$^{1,11}$ , 
  M.~\textsc{Feijoo-Fontan}$^{6}$ , 
  H.~\textsc{Fujioka}$^{12}$ , 
  Y.~\textsc{Gao}$^{11,13,14}$ , 
  H.~\textsc{Geissel}$^{4,9}$ , 
  F.~\textsc{Goldenbaum}$^{15}$ , 
  A.~\textsc{Gra\~na Gonzalez}$^{6}$ , 
  E.~\textsc{Haettner}$^{4}$ , 
  M. N.~\textsc{Harakeh}$^{10}$ , 
  Y.~\textsc{He}$^{11,14}$ , 
  H.~\textsc{Heggen}$^{4}$ , 
  C.~\textsc{Hornung}$^{4}$ , 
  N.~\textsc{Hubbard}$^{4,16}$ , 
  K.~\textsc{Itahashi}$^{17,18}$ , 
  M.~\textsc{Iwasaki}$^{17,18}$ , 
  N.~\textsc{Kalantar-Nayestanaki}$^{10}$ , 
  A.~\textsc{Kasagi}$^{11,19}$ , 
  M.~\textsc{Kavatsyuk}$^{10}$ , 
  E.~\textsc{Kazantseva}$^{4}$ , 
  A.~\textsc{Khreptak}$^{20,21}$ , 
  B.~\textsc{Kindler}$^{4}$ , 
  R.~\textsc{Knoebel}$^{4}$ , 
  H.~\textsc{Kollmus}$^{4}$ , 
  D.~\textsc{Kostyleva}$^{4}$ , 
  S.~\textsc{Kraft-Bermuth}$^{22}$ , 
  N.~\textsc{Kurz}$^{4}$ , 
  E.~\textsc{Liu}$^{11,13,14}$ , 
  B.~\textsc{Lommel}$^{4}$ , 
  V.~\textsc{Metag}$^{9}$ , 
  S.~\textsc{Minami}$^{4}$ , 
  D. J.~\textsc{Morrissey}$^{23}$ , 
  P.~\textsc{Moskal}$^{21,24}$ , 
  I.~\textsc{Mukha}$^{4}$ , 
  A.~\textsc{Muneem}$^{11,25}$ , 
  M.~\textsc{Nakagawa}$^{11}$ , 
  K.~\textsc{Nakazawa}$^{19}$ , 
  C.~\textsc{Nociforo}$^{4}$ , 
  H. J.~\textsc{Ong}$^{13,26,27}$ , 
  S.~\textsc{Pietri}$^{4}$ , 
  J.~\textsc{Pochodzalla}$^{2,3}$ , 
  S.~\textsc{Purushothaman}$^{4}$ , 
  E.~\textsc{Rocco}$^{4}$ , 
  J. L.~\textsc{Rodr\'iguez-S\'anchez}$^{6,28}$ , 
  P.~\textsc{Roy}$^{4}$ , 
  R.~\textsc{Ruber}$^{29}$ , 
  T.R.~\textsc{Saito}$^{4,11,14}$ , 
  S.~\textsc{Schadmand}$^{4}$ , 
  C.~\textsc{Scheidenberger}$^{4,9}$ , 
  P.~\textsc{Schwarz}$^{4}$ , 
  R.~\textsc{Sekiya}$^{17,18,30}$ , 
  V.~\textsc{Serdyuk}$^{15}$ , 
  M.~\textsc{Skurzok}$^{21,24}$ , 
  B.~\textsc{Streicher}$^{4}$ , 
  K.~\textsc{Suzuki}$^{4,31}$ , 
  B.~\textsc{Szczepanczyk}$^{4}$ , 
  Y. K.~\textsc{Tanaka}$^{11}$ , 
  X.~\textsc{Tang}$^{13}$ , 
  N.~\textsc{Tortorelli}$^{4}$ , 
  M.~\textsc{Vencelj}$^{7}$ , 
  H.~\textsc{Wang}$^{11}$ , 
  T.~\textsc{Weber}$^{4}$ , 
  H.~\textsc{Weick}$^{4}$ , 
  M.~\textsc{Will}$^{4}$ , 
  K.~\textsc{Wimmer}$^{4}$ , 
  A.~\textsc{Yamamoto}$^{32}$ , 
  A.~\textsc{Yanai}$^{11,33}$ , 
  J.~\textsc{Yoshida}$^{11,34}$ , 
  J.~\textsc{Zhao}$^{4,35}$ , 
  ~\textsc{for the WASA-FRS and Super-FRS Experiment Collaboration}
}

\inst{
  $^1$Instituto de Estructura de la Materia - CSIC, Madrid, Spain \\ 
  $^2$Institute for Nuclear Physics, Johannes Gutenberg University, Mainz, Germany,				   \\ 
  $^3$Helmholtz Institute Mainz, Johannes Gutenberg University, Mainz, Germany,					   \\ 
  $^4$GSI Helmholtzzentrum f\"ur Schwerionenforschung GmbH, Darmstadt, Germany,					   \\ 
  $^5$Institut f\"ur Kernphysik, Universit\"at K\"oln, Koln, Germany,						   \\ 
  $^6$IGFAE, Universidade de Santiago de Compostela, Santiago de Compostela, Spain,					   \\ 
  $^7$Jozef Stefan Institute, Ljubljana, Slovenia,								   \\ 
  $^8$University of Ljubljana, Ljubljana, Slovenia,								   \\ 
  $^9$Justus-Liebig Universit\"at Giessen, Giessen, Germany,									   \\ 
  $^{10}$University of Groningen, Groningen, The Netherlands,							   \\ 
  $^{11}$High Energy Nuclear Physics Laboratory, RIKEN Cluster for Pioneering Research, RIKEN, Wako, Japan, \\
  $^{12}$Tokyo Institute of Technology, Tokyo, Japan,								   \\ 
  $^{13}$Institute of Modern Physics, Chinese Academy of Sciences, Lanzhou, China,				   \\ 
  $^{14}$School of Nuclear Science and Technology, University of Chinese Academy of Sciences, Beijing, China,	   \\ 
  $^{15}$Institut fur Kernphysik, Forschungszentrum Julich, Julich, Germany,					   \\ 
  $^{16}$Institut fur Kernphysik, Technische Universitat Darmstadt, Darmstadt, Germany,				   \\ 
  $^{17}$Meson Science Laboratory, Cluster for Pioneering Research, RIKEN, Wako, Japan,	   \\ 
  $^{18}$Nishina Center for Accelerator-Based Science, RIKEN, Wako, Japan,		   \\ 
  $^{19}$Graduate School of Engineering, Gifu University, Gifu, Japan ,						   \\ 
  $^{20}$INFN, Laboratori Nazionali di Frascati, Frascati, Roma, Italy,						   \\ 
  $^{21}$Institute of Physics, Jagiellonian University, Krakow, Poland,						   \\ 
  $^{22}$TH Mittelhessen University of Applied Sciences, Giessen, Germany,						   \\ 
  $^{23}$National Superconducting Cyclotron Laboratory, Michigan State University, East Lansing, USA,		   \\ 
  $^{24}$Center for Theranostics, Jagiellonian University, Krakow, Poland,					   \\ 
  $^{25}$Faculty of Engineering Sciences, Ghulam Ishaq Khan Institute of Engineering Sciences and Technology, Topi, Pakistan, \\ 
  $^{26}$Joint Department for Nuclear Physics, Lanzhou University and Institute of Modern Physics, Lanzhou, China, \\ 
  $^{27}$Research Center for Nuclear Physics, Osaka University, Osaka, Japan,					   \\ 

  $^{28}$CITENI, Universidade da Coru\~{n}a, Ferrol, Spain, \\ 
  $^{29}$Uppsala University, Uppsala, Sweden,									   \\ 
  $^{30}$Kyoto University, Kyoto, Japan,									   \\ 
  $^{31}$Ruhr-Universit\"at Bochum, Institut f\"ur Experimentalphysik I, Bochum, Germany,				   \\ 
  $^{32}$KEK, Tsukuba, Japan,										   \\ 
  $^{33}$Saitama University, Sakura-ku, Japan,								   \\ 
  $^{34}$Tohoku University, Sendai, Japan,									   \\ 
  $^{35}$Peking University, Beijing, China,									 \\ 
}

\email{christophe.rappold@csic.es}

\recdate{March 2, 2026}

\abst{
The current understanding of light hypernuclei, subatomic nuclei exhibiting strangeness, is being actively investigated by several collaborations. Recent studies employing high-energy heavy-ion beams have revealed unexpected results for the three-body bound state $^3_\Lambda$H, known as the hypertriton. Experimental measurements of its lifetime and binding energy across multiple experiments have produced a puzzling situation referred to as the ``hypertriton puzzle''. This puzzle is the central focus of the data analysis from the WASA-FRS HypHI experiment at GSI-FAIR within the SuperFRS Experiment collaboration. Light hypernuclei are being investigated using 1.96 GeV/u $^6$Li beams impinging on a fixed carbon target, in conjunction with the WASA detector system and the Fragment Separator (FRS) at GSI-FAIR. The WASA-FRS experimental campaign was carried out during the first quarter of 2022. The primary objective of the experiment is to measure the lifetimes of $^3_\Lambda$H and $^4_\Lambda$H. The possible observation of the nn$\Lambda$ state constitutes the second component of the WASA-FRS research program. This contribution provides an overview of our investigation of hypernuclei. Focusing on the WASA-FRS HypHI experiment, details of the current analysis status are presented.

}

\kword{heavy-ion reaction, hypernuclei, lifetime}

\begin{document}
\maketitle

\section{Introduction}

In contemporary hypernuclear physics, the light sector around $^3_\Lambda$H, $^4_\Lambda$H, $^4_\Lambda$He, and nn$\Lambda$ has been the focus of several experimental and theoretical studies in recent years. Bound states of nucleons and hyperons, termed hypernuclei, serve as case studies of the baryon-baryon interaction within the SU(3)-flavor symmetry framework used to model baryonic matter from low to high densities. They provide a primary means to extract nucleon–hyperon and hyperon–hyperon interaction strengths, which were inaccessible prior to the advent of femtoscopy at the LHC and RHIC colliders \cite{femto1-alice,femto-star} and hadronic scattering experiments at J-PARC \cite{scat-jparc1,scat-jparc2}. The structure of the lightest hypernucleus, the hypertriton ($^3_\Lambda$H), remains under active investigation by multiple collaborations. In recent years, new results have been presented and published on lifetime estimates \cite{lt-h3l1,lt-h3l2,lt-h3l3} and the $\Lambda$ binding energy \cite{bL-1,bL-2,lt-h3l2,bL-3}. 

In this context, the WASA-FRS/Super-FRS collaboration planned and executed the WASA-FRS campaign in 2022 under the FAIR-Phase 0 program at the GSI–FAIR facility. Two experiments were conducted: one focusing on the spectroscopy of $\eta^\prime$-mesic nuclei in proton-induced reactions \cite{etaprime-wasa}, and a second centered on the spectroscopy of light hypernuclei in ion-induced reactions (HypHI) \cite{hyp-wasa}. The hypernuclear study aimed to reproduce the experimental conditions of the 2009 HypHI Phase 0 experiment with a distinct setup. The objectives were to enhance spectroscopic precision to address the hypertriton puzzle by providing more accurate lifetime estimates and to test the existence of the nn$\Lambda$ bound state. The WASA-FRS HypHI experiment again employed a $^6$Li beam at 1.96 GeV/u impinging on a diamond-carbon target. The Phase 0 experiment had demonstrated that hypernuclear spectroscopy can be performed via ion-induced reactions \cite{hyp-p0-1,hyp-p0-2}.

Hypernuclear production is explained by the coalescence of hyperons produced in the participant zone with the spectator flying out of the target due to the Lorentz boost at relativistic beam energy. Coalescence is enabled by the broad rapidity distribution of the produced hyperons, which overlaps with the spectator phase space. Hypernuclear production via this mechanism can thus probe the temporal overlap between participants and spectators and, consequently, the collision dynamics. Hypernuclei can also be formed in the participant zone, as first demonstrated by STAR \cite{STAR-hyp} and subsequently by ALICE \cite{ALICE-hyp} and HADES \cite{HADES-hyp}, at the freeze-out of the quark–gluon plasma or hadron gas produced in the collisions. In the spectator region, secondary ($\pi$, K) and (K, $\pi$) processes may also form hypernuclei, albeit with much smaller probability \cite{hyp-pi}.

In the HypHI Phase 0 experiment, hypernuclei were observed via in-flight spectroscopy by reconstructing the invariant mass of their mesonic weak decays \cite{hyp-p0-1}. The Lorentz boost of the projectile-spectator fragments allowed the produced hypernuclei to exit the target material and decay in the laboratory vacuum. Hypernuclei were identified through their two-body $\pi^-$ decays, reconstructed from track measurements across a magnetic dipole and particle-tracking detectors. These results constituted important milestones for the HypHI collaboration and established the experimental concept as a novel tool for producing and studying hypernuclei. However, the precision of the lifetime and invariant-mass observables was limited by the experimental setup. The possible observation of nn$\Lambda$ \cite{hyp-nnL} has been an intriguing puzzle, as several theoretical calculations indicate that such a state should not be bound \cite{th-nnL1,th-nnL2,th-nnL3,th-nnL4,th-nnL5,th-nnL6}. To improve upon the previous observations, a new experimental setup was designed to address the limitations of the earlier experiment. This upgraded setup, the WASA-FRS HypHI experiment, aims to contribute to resolving the outstanding questions surrounding three-body hypernuclear states.

\section{WASA-FRS experiment}

\begin{figure}[htb]
  \centering
  \begin{tabular}{cc}
    \resizebox{80mm}{!}{\includegraphics{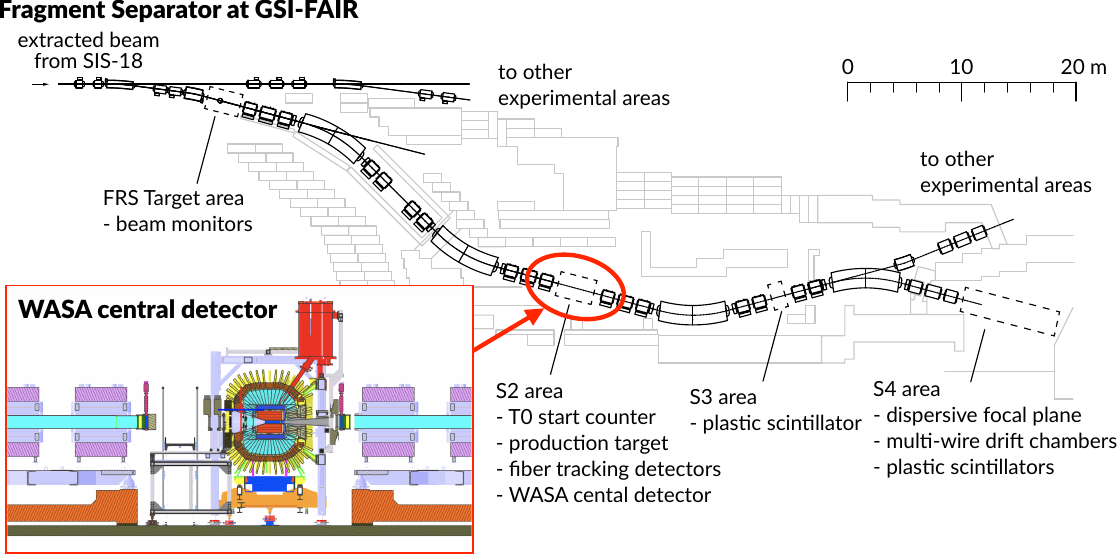}}
    & 
    \resizebox{70mm}{!}{\includegraphics{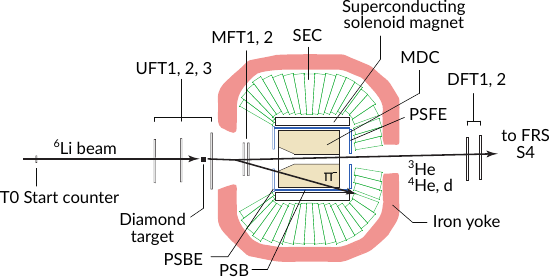}} 
  \end{tabular}
    \caption{WASA-FRS experimental setup. Left panel: layout of the FRagment Separator at GSI-FAIR, where the beam is directed toward the S2 area, where the WASA detector apparatus was located and the production of hypernuclei took place. The FRS segment from S2 to S4 serves as a high-resolution forward spectrometer for the decay fragments. Right panel: the cylindrical WASA detector apparatus located in the S2 area, which is used to measure light particles, such as pions, within the superconducting solenoid magnet.}
  \label{fig:layout}
\end{figure}

The WASA-FRS experimental setup consists of the WASA central detection apparatus and the FRagment Separator (FRS) of the GSI-FAIR facility. The WASA central detector system was previously located at COSY in Jülich, Germany \cite{wasa-cosy1, wasa-cosy2}, and was transported to GSI-FAIR in 2019. The WASA system was refurbished with a new electronics readout system, a new scintillator barrel and walls, and a new cryogenics system for the superconducting solenoid magnet. The WASA electronics readout was also developed to integrate into the GSI-FAIR data acquisition and trigger system. WASA was placed at the mid-focal plane of the FRS separator, where the hypernuclear production target was located. It is responsible for measuring the hadrons produced in the nuclear collisions, while the second half the of FRS is employed as a high-resolution spectrometer for the forward decay fragments. In an exclusive experiment, the mesonic weak decays $^3_\Lambda$H $\to$ $^3$He+$\pi^-$, $^4_\Lambda$H $\to$ $^4$He+$\pi^-$ and nn$\Lambda \to$d+n+$\pi^-$ were the focus in order to assess the feasibility of the experimental method and to contribute to resolving the puzzles described previously. In these decays, the $\pi^-$ meson is measured by the WASA apparatus, while a narrow magnetic-rigidity acceptance window is set for the FRS to measure precisely the momentum of the outgoing decay fragment ($^3$He, $^4$He, d). As shown in Fig. \ref{fig:layout}, the combination of both detection systems is employed to obtain a more precise momentum measurement of the decay products thereby improving the invariant-mass reconstruction of the hypernuclei.

The magnetic field of the WASA solenoid magnet was set to1T. WASA included a calorimeter with CsI crystals (SEC), an inner drift chamber with straw tubes (MDC), and newly developed arrays of plastic barrel hodoscopes (PSB) \cite{nima-psb1,nima-psb2} and end-cap hodoscopes (PSFE and PSBE), as shown in the right panel of Fig.~\ref{fig:layout}. In addition to the WASA apparatus, new scintillating fiber detectors were installed. Two fiber detector stations, UFT1 and UFT2, were placed in front of the diamond target to precisely track the incoming beam. Immediately behind the target, another fiber station, UFT3, was placed in front of the WASA detectors for the determination of the interaction point of the collisions. Two fiber detector stations, DFT1 and DFT2, were positioned behind the WASA detector to track the outgoing fragments before they entered the S2-S4 section of the FRS. Furthermore, a mini-fiber detector comprising two fiber stations, MFT1 and MFT2, was situated inside the iron yoke of the WASA solenoid magnet. Additionally, a small hodoscope composed of plastic scintillator fingers, referred to as the T0 start counter \cite{nima-T0}, was mounted in front of UFT1 to characterize the beam particles.

A diamond target with a thickness of 9.87 g/cm$^2$ was located between the tracking stations UFT2 and UFT3, where hypernuclear production occurred during collisions of a $^6$Li beam at 1.96 $A$GeV with the $^{12}$C target. $\pi^-$ mesons from hypernuclear decays were measured by the WASA detectors together with the stations MFT1 and MFT2. Residual nuclear fragments from hypernuclear decays were transported to and analyzed in the S2–S4 sections of the FRS. Details of the experimental concept and simulated expectations were presented in \cite{hyp-wasa}. The acceptance and efficiencies were evaluated, and the invariant-mass resolution was expected to be at least twice as good, with at least an order of magnitude more statistics than were collected during the first HypHI experiment.

\section{Preliminary results}

To identify fragments transmitted through the FRS beyond the S2 focal plane, time-of-flight and energy-loss measurements from scintillators at the S3 and S4 focal planes enabled an unambiguous identification of $^3$He, $^4$He, and deuterons. Momentum measurement and analysis in the S2–S4 section of the FRS were performed via ion-optics calibration, yielding a momentum-reconstruction resolution of $5 \times 10^{-4}$. Overall, fragment identification and analysis performance in the FRS were satisfactory.

\begin{figure}[htb]
  \centering
  \begin{tabular}{c}
    \resizebox{!}{40mm}{\includegraphics{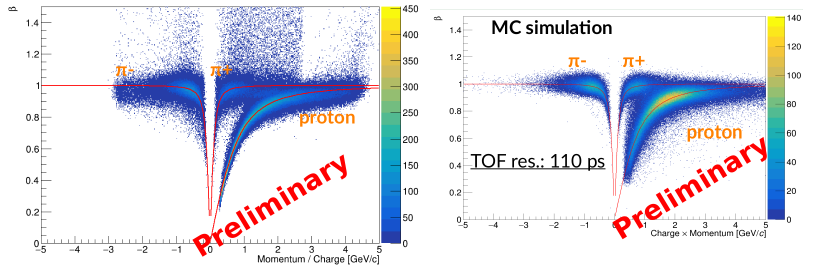}} 
  \end{tabular}
    \caption{Particle identification within the WASA central detection system. Owing to the $\beta$–momentum correlation, light hadrons such as $\pi^-$, $\pi^+$, and protons can be identified. The left and right panel show the results obtained from the experimental data and the Monte Carlo simulation, respectively. In the Monte Carlo simulation, a time-of-flight resolution of 110 ps is considered.}
  \label{fig:pid}
\end{figure}

At S2, the reconstruction of hadrons produced in WASA was performed using a Graph Neural Network (GNN) \cite{gnn-wasa}. The GNN was trained to leverage hit information from the tracking detectors and hodoscopes to find particle tracks and estimate their track parameters. The final GNN architecture comprised 5 final heads: a particle classifier for identified tracks, one head for each particle's full momentum estimation (3 heads in total), and a combined head for position and track length estimation. This configuration enabled highly efficient track reconstruction, with a momentum resolution of 8.8\%, and an angular resolution of 2.3 mrad. Fig. \ref{fig:pid} presents particle identification results obtained from both experimental data and Monte Carlo simulations. The simulations were based on a Geant4 model of the S2 experimental setup. Detector position calibrations and the position and time resolutions estimated during the experiment were propagated into the simulations. A good correspondence between the two particle identification plots is observed.

\begin{figure}[htb]
  \centering
  \begin{tabular}{cc}
    \resizebox{!}{40mm}{
    \begin{overpic}[scale=0.4]
      {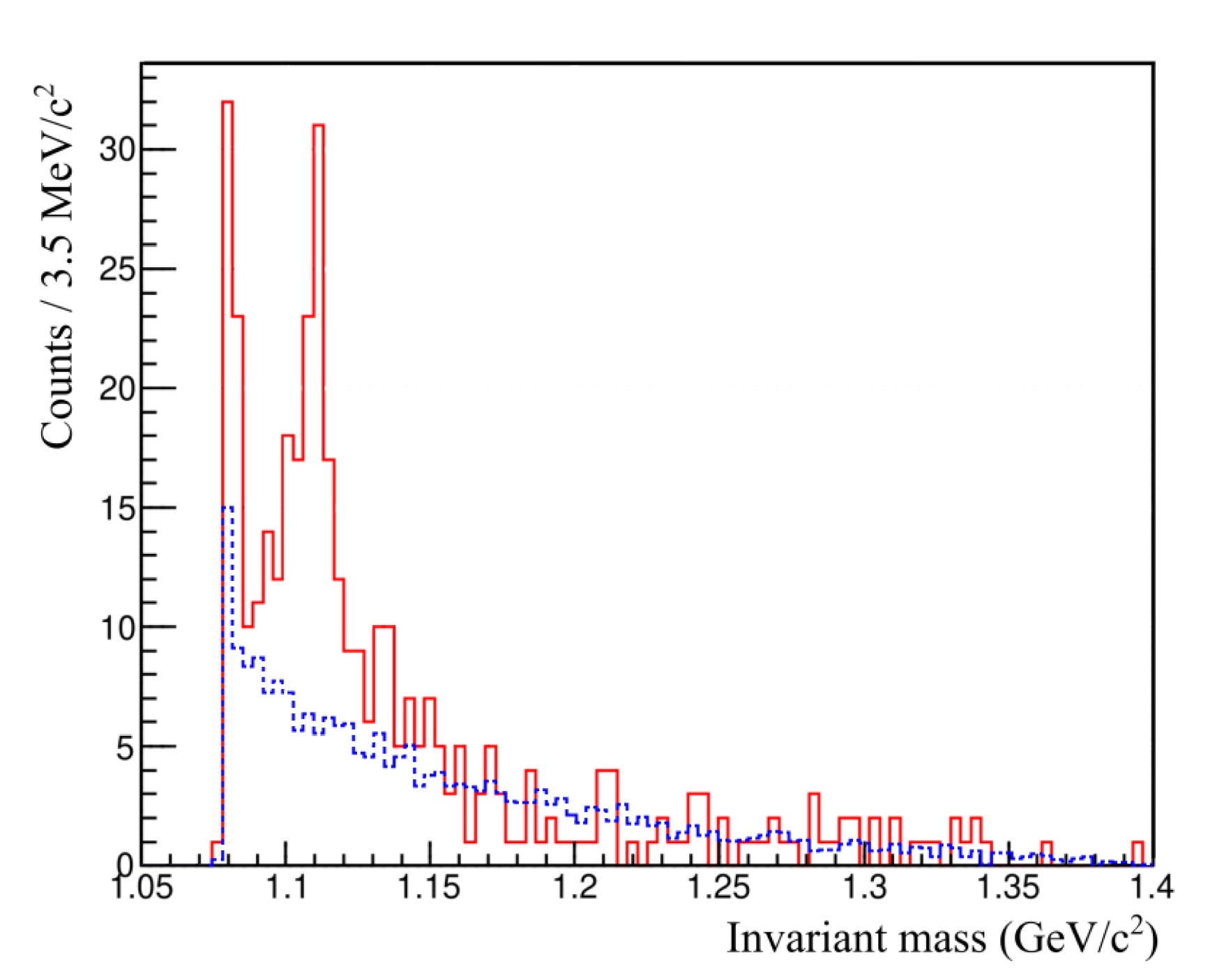}
       \put(180,130){\rotatebox{35}{\Huge{\textbf{\color{red}Preliminary}}}}

    \end{overpic}
    }

    &
      \resizebox{!}{40mm}{
      \begin{overpic}[scale=0.8]
        {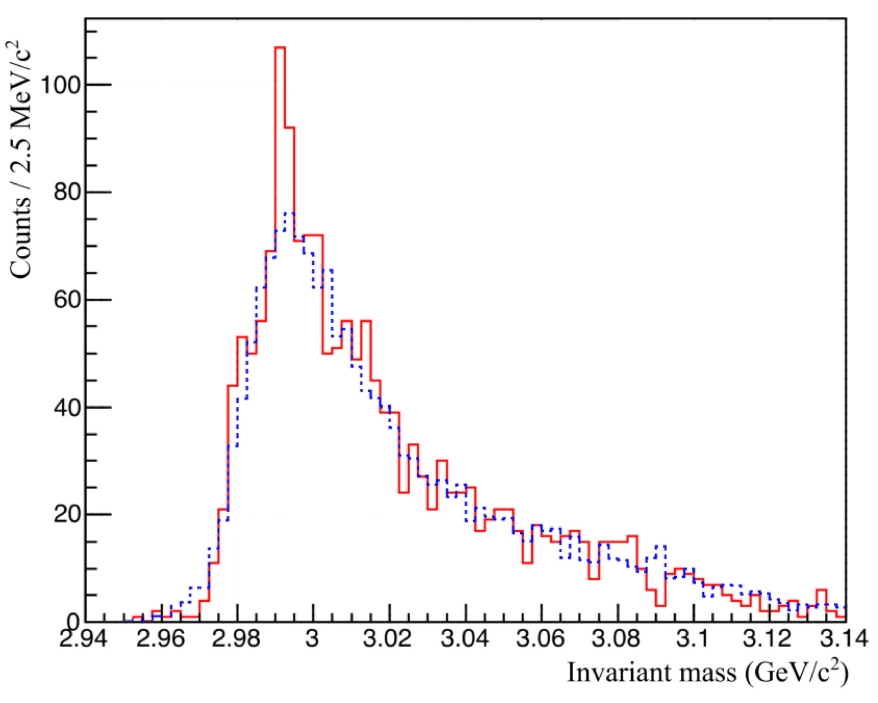}
        \put(190,130){\rotatebox{35}{\Huge{\textbf{\color{red}Preliminary}}}}
      \end{overpic}
      }
      
  \end{tabular}
    \caption{Invariant mass spectra of the $\Lambda$ hyperon and $^3_\Lambda$H, shown in the left and right panels, respectively. The vertex selection of 15 cm downstream of the target was applied. The red distribution corresponds to the experimental data, while the dashed blue distribution represents the results of the mixed-event analysis, which combines independent pairs of pions and fragments originating from distinct recorded events. Signals of the $\Lambda$ hyperon and $^3_\Lambda$H are observed above the mixed-event distribution.}
  \label{fig:invmass}
\end{figure}

Thanks to robust track reconstruction in WASA and the precise measurement of decay fragments between the S2–S4 segment of the FRS, event reconstruction can be performed. When an identified $\pi^-$ is measured in coincidence with a well-measured $^3$He fragment, a secondary vertex can be reconstructed. Because the measured $^3$He fragments mainly originate from fragmentation reactions in nuclear collisions, the expected signal-to-background ratio of the hypernuclear signal is approximately $2 \times 10^{-4}$. To improve the signal-to-background ratio, several selection criteria must be applied to the secondary-vertex candidates. By requiring a longitudinal position along the beam axis of 15 cm, the invariant-mass distribution of the ($\pi^-$, $^3$He) pair can be observed, as shown in red in the right panel of Fig. \ref{fig:invmass}. To cross-check the reconstruction and the performance of the WASA detection system, a similar reconstruction for the $\Lambda \to \pi^- + p$ decay was performed. The difference is that the protons are measured inside the WASA apparatus and not in the FRS. The result of the vertex reconstruction can be seen in the invariant mass distribution in the left panel of Fig. \ref{fig:invmass} in red. A mixed-event analysis was also performed to estimate the background-only contribution and is shown in dashed blue in the respective invariant-mass spectra of Fig. \ref{fig:invmass}. The mixed-event analysis consists of taking uncorrelated pairs of decay products from distinct events to ensure background-only invariant mass distribution. Compared with the mixed-event distribution, only a limited amount of genuine signal is observed. These results correspond to our analysis efforts during 2024. Based on the preliminary observation of a hypernuclear signal, efforts in 2025 have focused on improving the signal-to-background ratio to obtain a clearer hypernuclear signal, enabling the extraction of the hypernuclear lifetime. Using a machine-learning method based on the XGBoost library, a supervised classifier was developed to separate the hypernuclear signal from all background sources.

\begin{figure}[htb]
  \centering
  \begin{tabular}{cc}
    \resizebox{!}{40mm}{\includegraphics{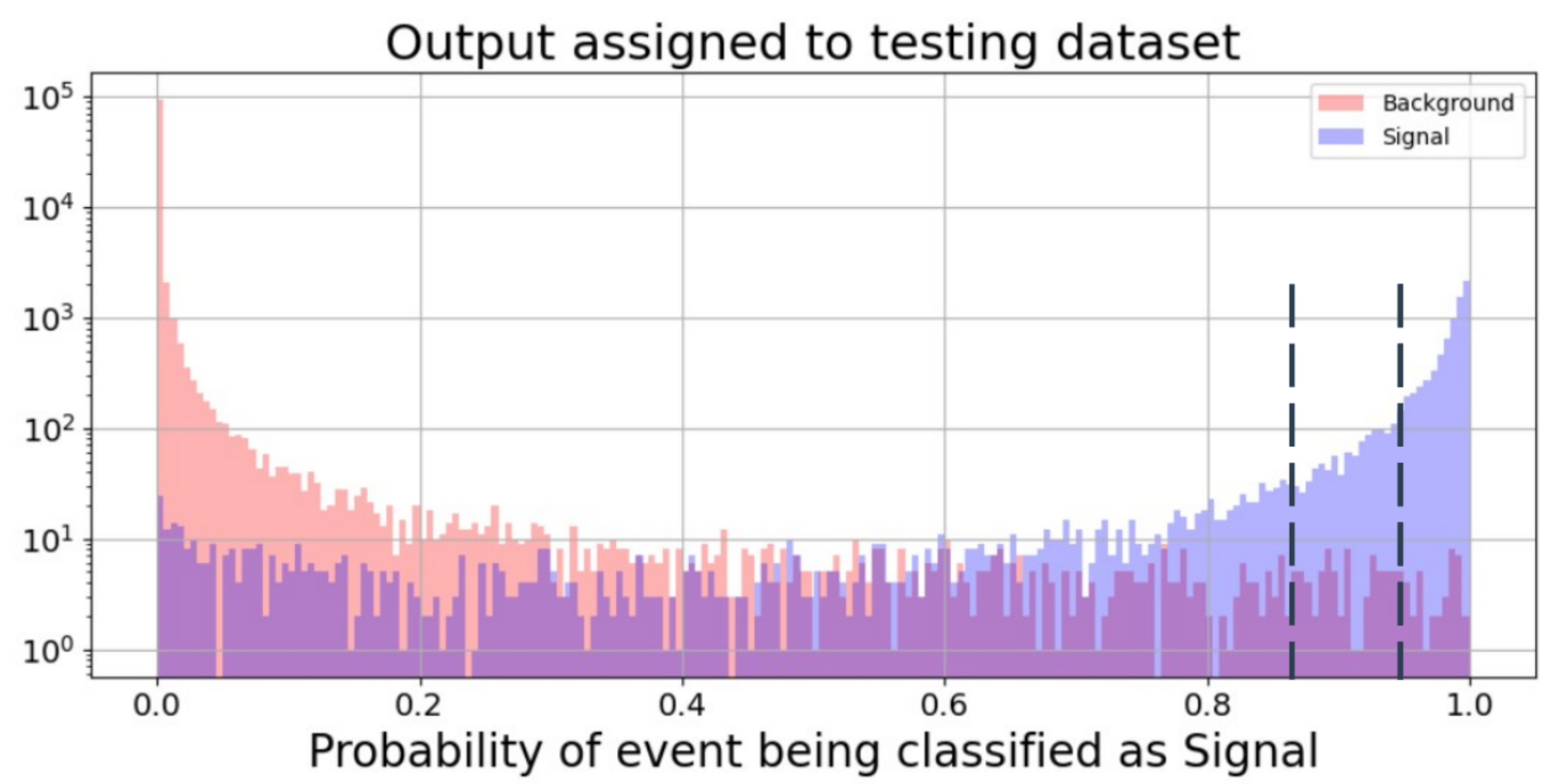}} 
    &
      \resizebox{!}{40mm}{\includegraphics{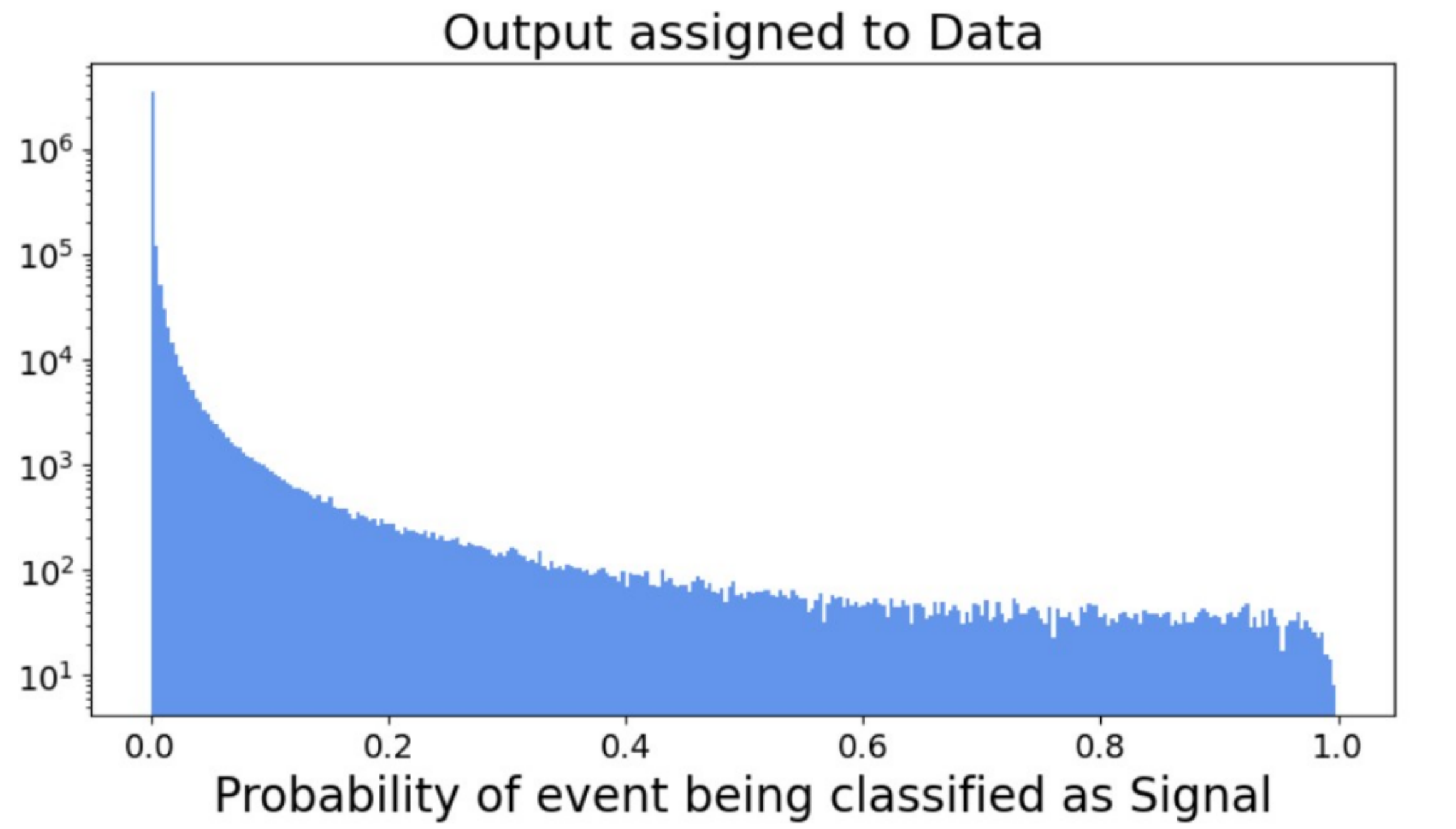}} 

  \end{tabular}
    \caption{Output of the Boosted Decision Tree model classifying events into signal and background classes. The left panel shows the output of the optimal machine learning model after training when applied to the test dataset, demonstrating a clear separation between the two classes. The right panel illustrates the application of the model to the full experimental dataset. With a high threshold on the model output, a hypernuclear signal with high purity is obtained.}
  \label{fig:bdt}
\end{figure}

The features used as input to the decision tree model are physical observables related to the geometry of the decay vertex, the hit pattern on the fiber trackers, and the $\pi^-$ hit multiplicity on the MFT fiber trackers and the MDC detectors. An important decision was made to exclude the invariant mass and lifetime observables as training features in order to obtain a signal-background classification uncorrelated with these observables and thereby avoid bias in the signal selection. In Fig. \ref{fig:bdt}, the output of the optimal machine learning model obtained after training is displayed.

In Fig. \ref{fig:bdt}, the left panel shows the application of the trained model to the test dataset that was not used during training, demonstrating the efficiency and purity of the model in discriminating signal from background: with a threshold value of 0.85, the model is able to select hypernuclear signal with high purity, for which a very small background contribution is expected. The right panel of Fig. \ref{fig:bdt} shows the application of the optimal model to the $^3_\Lambda$H experimental dataset. With a threshold value from 0.85 to 0.95 as indicated by the vertical lines in the left panel, several hunderds of hypertriton signal is observed. In evaluation, the signal-to-background ratio was 2.4, corresponding to a improvement factor of 12000. After signal selection based on this machine learning method, the invariant mass distribution of $^3_\Lambda$H was examined, and a clear signal was observed with hundreds of $^3_\Lambda$H candidates. The experimental lifetime was then inferred. A lifetime estimate was obtained with a statistical uncertainty of approximately 30 ps and a systematic uncertainty of approximately 15 ps. The distribution and extracted values will be reported promptly in a dedicated article and will not be reported in this conference proceeding.

The same analysis procedure was applied to $^4_\Lambda$H and the $\Lambda$ hyperon, and good results were also observed, with clean invariant-mass distributions, high signal purity, and extracted lifetimes comparable to the literature with precise uncertainties. A cross-check analysis was also performed using a mixed-event dataset, in which no hypernuclear signal is present. A model trained with this dataset as signal did not generate any invariant-mass spectra compatible with a hypernuclear signal, and the extracted lifetime was 71 $\pm$ 4 ps, all of these distributions are compatible with an analysis of the background.

\section{Perspective}

With the application of machine learning to the classification of hypernuclear signals, a clean observation of $^3_\Lambda$H and $^4_\Lambda$H hypernuclei was achieved. This yielded a clean dataset of hypernuclear decays, enabling an improved lifetime estimation compared with the HypHI Phase 0 experiment. An order of magnitude more clean hypernuclear events was obtained, and the extracted lifetime uncertainty was reduced by a factor of 2. It can be asserted that the WASA-FRS HypHI experiment achieved its primary goal of improving upon the Phase 0 experiment. Owing to the clean dataset of hypernuclear decays, other physical observables will be extracted, such as the hypernuclear radius, using the procedure developed in \cite{radius-wasa}.

\section{Acknowledgement}

The two WASA-FRS experiments, S447 and S490 were conducted in 2022 at the FRS of the GSI Helmholtzzentrum fuer Schwerionenforschung, Darmstadt (Germany) within the framework of FAIR Phase 0. The authors would like to acknowledge the GSI staff for their support during the experiments. The WASA-FRS project is supported by JSPS KAKENHI (Grant No. JP18H01242), the JSPS Grant-in-Aid for Early-Career Scientists (Grant No. JP20K14499), and the JSPS Fostering Joint International Research (B) (Grant No. JP20KK0070); Proyectos I+D+i 2020 ref: PID2020-118009GA-I00; Proyecto Consolidación Investigadora 2022 ref: CNS2022-135768, Proyectos I+D+i 2022; ref: PID2022-140162NB-I00; grants 2019-T1/TIC-13194 and 2023-5A/TIC-28925 of the program Atracción de Talento Investigador of the Community of Madrid, ``Programa Momentum-CSIC'', ref: MMT24-IEM-02, the Regional Government of Galicia under the Postdoctoral Fellowship Grant No. ED481D-2021-018, the SciMat and qLife Priority Research Areas budget under the program Excellence Initiative - Research University at the Jagiellonian University and the European Unions Horizon 2020 research and innovation programme under grant agreement No. 824093. The project is also supported by the Justus-Liebig-Universität (JLU), Gießen, Germany, and GSI under the JLU-GSI strategic Helmholtz partnership agreement.


\begin{thebibliography}{40}
\bibitem{femto1-alice} ALICE Collaboration, Nature \textbf{588}, 232 (2020). 
\bibitem{femto-star} STAR Collaboration, arXiv:2511.15493 (2025)
\bibitem{scat-jparc1}  K. Miwa et al., Phys. Rev. C, \textbf{104}, 045204 (2021).
\bibitem{scat-jparc2}   K. Miwa et al., Phys. Rev. Lett., \textbf{128}, 072501 (2022).
\bibitem{lt-h3l1}     STAR Collaboration, Nature \textbf{632}, 1026 (2024)
\bibitem{lt-h3l2} ALICE Collaboration, Phys. Rev. Lett., \textbf{131}, 102302 (2023)
\bibitem{lt-h3l3} STAR Collaboration, Phys. Rev. Lett., \textbf{128}, 202301 (2022)
\bibitem{bL-1} A. Kasagi et al., PTEP \textbf{2025}, 083D01 (2025)
\bibitem{bL-2} T. Akaishi, et al., Phys. Lett. B \textbf{873}, 140163 (2026) 
\bibitem{bL-3} STAR Collaboration, Nature Phys. \textbf{16}, 409 (2020)
\bibitem{etaprime-wasa} R.~Sekiya et al., Nuovo Cim.C \textbf{47} 230 (2024)
\bibitem{hyp-wasa} T. R. Saito et al.,     Nature  Rev. Phys. \textbf{3} 803 (2021)
\bibitem{hyp-p0-1} C. Rappold et al., Nucl.Phys. A \textbf{913} 170 (2013)
\bibitem{hyp-p0-2} C. Rappold et al., Phys. Lett. B \textbf{747} 129 (2015)
\bibitem{STAR-hyp} The STAR Collaboration, Science \textbf{328}, 58 (2010).
\bibitem{ALICE-hyp} J. Adam, et al., for ALICE collaboration, Phys. Lett. B \textbf{754}, 360 (2016)
\bibitem{HADES-hyp} R. Abou Yassine et al., arXiv:2512.12454 (2025)
\bibitem{hyp-pi} T.~Gaitanos, H.~Lenske and U.~Mosel. Prog. Part. Nucl. Phys. \textbf{62}, 439 (2009).
\bibitem{hyp-nnL} C. Rappold et al., Phys.Rev. C \textbf{88} 041001 (2013)
\bibitem{th-nnL1} E. Hiyama, et al., Phys. Rev. C \textbf{89}, 061302 (2014)
\bibitem{th-nnL2} A. Gal, et al., Phys. Lett. B \textbf{736}, 93 (2014)
\bibitem{th-nnL3} H. Garcilazo, et al., Phys. Rev. C \textbf{89}, 057001 (2014)
\bibitem{th-nnL4} M. Schafer, et al., Phys. Lett. B \textbf{808}, 135614 (2020)
\bibitem{th-nnL5} S.-I. Ando, et al., Phys. Rev. C \textbf{92}, 024325 (2015)
\bibitem{th-nnL6} F. Hildenbrand, et al., Phys. Rev. C \textbf{100}, 034002 (2019)

\bibitem{wasa-cosy1} P. Adlarson, et al., Phys. Rev. C \textbf{102}, 044322 (2020).
\bibitem{wasa-cosy2} P. Adlarson, et al., Phys. Lett. B \textbf{802}, 135205 (2020).

\bibitem{nima-psb1} R. Sekiya et al., Nucl. Instruments Methods Phys. Res. A \textbf{1034}, 166745 (2022).
\bibitem{nima-psb2} Y.K. Tanaka et al., Nucl. Instruments Methods Phys. Res. A \textbf{1083} 171065 (2026)
\bibitem{nima-T0} E. Liu et al., Nucl. Instruments Methods Phys. Res. A \textbf{1064} 169384 (2024)
\bibitem{gnn-wasa} H.~Ekawa et al., Eur. Phys. J. A \textbf{59}, 103 (2023)

\bibitem{radius-wasa} Y.~Gao et al., Chin. Phys. C \textbf{49}, 124003 (2025)
  
\end{thebibliography}
\end{document}